
\magnification=\magstep1
\hsize 32 pc
\vsize 42 pc
\baselineskip=24 true pt
\centerline{\bf Synchronized First-Passages in a Double-Well
System}
\centerline{\bf Driven by Asymmetric Periodic Field}
\vskip 1.0 true cm
\centerline{Mangal C. Mahato and A.M. Jayannavar}
\vskip 0.3 true cm
\centerline{Institute of Physics}
\centerline{Sachivalaya Marg, Bhubaneswar - 751 005, INDIA}
\vskip 1.0 true cm
\noindent{\bf Abstract}
\vskip .3 true cm
	We perform Langevin dynamic numerical simulation on a
double-well potential system subjected to an asymmetric
saw-tooth type external time varying field and white noise
forces. The hysteresis loss calculated from the first-passage
time distribution obtained shows asymmetric behaviour with
respect to the asymmetry in the field sweep. The
hysteresis loss, in our model, being a measure of the
synchronized passages from one well to the other, indicates
asymmetric "correlated" passages in the two opposing directions
when driven by a temporally asymmetric external field in the
presence of white noise (fluctuating) forces. The implication of
our results on the phenomena of predominantly unidirectional
motion of a Brownian particle in a symmetric periodic
(nonratchet-like) potential is discussed.
\vfill
\eject
\vskip .3 true cm
	Recently, there has been some increased activity on the
asymmetric transport, without any obvious applied bias, in
periodic systems. Many physical models have been proposed for
possible explanations of such phenomenon.[1-5] These are theoretical
models and need not be the theories of asymmetric transport observed
   in nature(for example, motion of macromolecules along
   microtubules. Yet, they provide impetus to our present
understanding of transport phenomena in nonequilibrium systems.
Understandably, these models incorporate the ideas of
nonlinearities and stochasticities involved; the ingredients of
the models in most of the cases are: the nature of periodic
potential (ratchet- or nonratchet-like, time independent or
fluctuating, etc.), the driving field, and the fluctuating
force(s). In the present work we explore possible asymmetric
passages from one well to the other in a symmetric two-well
system under the influence of additive white noise fluctuating
forces when the system is externally driven by a temporally
asymmetric but a zero-time averaged field. It is to be noted,
however, that the effect of the temporally asymmetric field has
already been explored,[6] but mostly for deterministic systems.
Also, very recently, the role of temporally asymmetric
fluctuations towards the operation of correlation ratchets has
been investigated.[7] We, in this work, do arrive at the results
not directly but through the study of one of the very important
properties of nonequilibrium systems, namely, the hysteresis.

	Hysteresis is an important property of inhomogeneous
macroscopic systems. These inhomogeneities could be due to the
presence of first-order phase boundaries, domain-walls, bulk,
surface, or point defects, etc., and inhibit the process of
fluctuation-assisted homogenization, in the time scales of
experimental  observation, of the macroscopic system in the
absence of externally applied fields. The response of the system
parameters characterizing the state of the macroscopic system to
external fields generally involves time delays resulting in the
accumulation and subsequent sudden release of strains leading to
frictional losses. The frictional losses cause irreversibility
with respect to the response to external field reversals. This
irreversibility or the hysteretic property of the macroscopic
system depends on the external field sweep rates in an important
way. In the last one decade or so there has been a good amount of work
on the kinetic aspect of hysteresis in various model systems
including the double-well potential system[8]; for example, in
Ref.8 the scaling behaviour oh hysteresis loss with respect to
the external field parameters is discussed and, moreover, the
hysteresis loss is shown to exhibit stochastic resonance
property with respect to the strength of the external noise
(fluctuations) . In the present work
we show that the hysteresis loss, which is a measure of the
noise-aided coherent passages across the potential barrier in a
two well system,[8] exhibits asymmetric behaviour with respect to
the temporal asymmetry of the external field sweep.

	In this article, we present a simple model to understand
synchronized motion of a particle in a two-well system
subjected to fluctuating(white noise) forces and to a
periodic (saw tooth type) external field,[Fig.1] with zero time-averaged
force, making the two wells move up and down synchronously.
A particle in one of the two minima, consequently, sees a
potential barrier for passage to the second well vary
synchronously to the external field. However, if we put the
particle in the second well
it too sees its barrier vary periodically, as the external field
is varied, but with a phase difference with respect to
the first. Thus if the variation of the external field is
symmetric (within a period) in time, for example a sinusoidal field
or a symmetric saw tooth type field, the net time averaged(over
a large number of cycles) particle flux across the potential
barrier will be zero, when the two (equally deep) wells
are taken to be equally populated in the beginning. We, now, ask
the question: What would happen if the external saw tooth field
is asymmetric, $\it i.e.$ the $\pm$slopes are different in
magnitude? One can easily see, from Fig.1, that, in this case too, each of
the particle will experience a zero-time-averaged externally
applied force-field. Yet, we expect a difference in the
synchronization behaviour of passage of the two particles.

	As mentioned earlier, the Brownian particle in a well
sees a potential barrier for its passage to the other well. The
potential barrier changes periodically as the external periodic
saw tooth field in time (Fig.1) is applied. The passage is expected to
take place mostly around the time when the potential barrier
height is the minimum depending on the fluctuating force
strength and how fast the external field is being changed, etc.
A passage may take many cycles of the external field sweep
before it is completed. An
appropriate choice of system parameters may, however,
yield synchronized(with respect to the applied field) passage of
the particle (Fig.2). This synchronized passage is linked to the
hysteretic behaviour of the two-well system.[8] The higher is the
passage synchronized the larger is the hysteresis loop area. Or
in other words, the larger is the hysteresis loop area the higher
is the efficiency of correlated passages. We thus perform,
following an earlier work on hysteresis[8], a numerical experiment
on the two-well (Landau) potential:
$$\Phi (m)=U(m)-h(t)m,\eqno{(1a)}$$
with the system potential field,
$$U(m)=-{a\over 2}m^2+{b\over 4}m^4,\eqno{(1b)}$$
where $m$ is the order parameter, for example, the displacement,
$a$ and $b$ are constants and $h(t)$ is the external time
dependent field.

	We proceed in the following way. We take $h(t)$ to have
a saw tooth behaviour, as shown in Fig. 1, with amplitude
$h_0<\mid h_c \mid $, where $h_c$ is the critical field at which
one of the two wells of the potential $(1a)$ disappears. The field
$h(t)$ peaks with value $h_0$ at $t=nT_0$, $n=0, 1, 2, ...$ and
acquires minimum value $-h_0$ at $t=(T_1+nT_0)$.  The asymmetry
of $\pm$slopes is obtained by taking $T_1\not= T_0/2$ and
measure as $\Delta ={(T_1-T_0/2)\over {T_0/2}}$.

	We prepare our system, at $t=0$, at the minimum of, say,
the right-side well when $h(0)=h_0$ (we always take $h_0
=0.7h_c$ so that the potential barrier never vanishes) and
let it evolve in time by the combined effect of the potential
$\Phi (m)$ (with $a=2$ and $b=1$) and a fluctuating
force $f(t)$. The evolution of the coordinate $m$ is described
by the overdamped Langevin equation
$$\dot m = -{\delta \Phi (m)\over {\delta m}}+f(t),\eqno(2)$$
with
$$<f(t)>=0,\eqno(3a)$$
and
$$<f(t)f(t')>=2D\delta (t-t').\eqno(3b)$$
The averaging $<...>$ is done over all possible
realizations of the random force $f(t)$. The experiment involves
recording the first passage times $\tau$ to the other (left)
well minimum of the system for the same initial conditions but
with different realizations of the random force $f(t)$[8]. The
records give the distribution $\rho (\tau)$ of $\tau$ (Fig.2). $\rho
(\tau)$ depends on various parameters including $D$, $h_0$,
$T_1$, $T_0$, etc. It may spread over many cycles of $h(t)$ and
generally peaks around $t=T_1+nT_0$, such that
$h(t)=h(T_1)=-h_0$. However, $\rho (\tau)$ can be converted into
$\rho (h_J)$, where $h_J$=$h(\tau)$. From $\rho (h_J)$ we
calculate[8] the upper half of the hysteresis loop $M(h_J)$:
$${M(h_J)\over h_c}=1-{2\over h_c}{\int _{h_J}^{h_0}
{\rho_(h_{J}^{'})} dh_{J}^{'}},\eqno{(4)}$$
with the saturation value of ${M\over h_c}=\pm 1$.
The lower half is completed by symmetry, meaning thereby that the
system is initally prepared at the minimum of the left well at
$h(t=0)=-h_0$ and drive the field in the reverse direction in
time to obtain $\rho (\tau)$ for passage to the right well, and
then calculate the
hysteresis loop area. The hysteresis loop area, as mentioned
earlier, is a measure of sychronized passages; for maximum
possible sychronization we have the rectangular hysteresis loop
with maximum area of $4h_0h_c$, corresponding to $\rho (\tau) =
{1\over N}\Sigma \delta (\tau -(nT_0 + T_1))$, the summation
being over $N\longrightarrow \infty$ set of $\tau$ calculations.

	The experiment is repeated for various values of
$\Delta$, the asymmetry in the field sweep rates, for given $D$,
$h_0$, etc. The plot of hysteresis loop area versus $\Delta$ is
shown in Fig. 3. The asymmetry of the hysteresis loop area about
$\Delta=0$ is quite revealing and becomes more pronounced for
lower $D$ values. The result shows that
for $\Delta <0$, passage from right well to the left is more
synchronized than for $\Delta >0$. With a little thought it can
be stated, in other words, as the passage from the right well to
the left well is more synchronized and hence more correlated
than the passage from left well to the right well for $\Delta
<0$, and vice versa.

	The external field sweep $h(t)$ satisfies, as it should
be clear from Fig.1, $\int
_{0}^{T_0} h(t) dt =0$ and the fluctuating force $f(t)$
satisfies eq.$(3a)$. The numerical experiment is performed under
two conditions, namely, the initial condition $m(t=0)=\bar
m_{2}(0)$, the minimum of the right well, for
example, and the condition that the particle gets absorbed as
soon as it reaches the left well minimum $\bar m_{1}(\tau)$.
The system is constructed to be dynamic and the time average of
 $\dot m$ in eq.(2) need not be non-zero to have an asymmetric
diffusive motion. This is not justified, however, for, say
$U(m)=\it {constant}$, or a frictionless flat potential surface
considering eq.(2). But, with a flat potential surface,
depending on the problem at hand, one would perhaps be wiser to
use the full Langevin equation rather than the overdamped eq.(2)
and, then, again it will be quite easy to verify that even
though the average velocity vanishes one can still have
asymmetric average displacement given the form of $h(t)$ with
$\Delta \not= 0$. However, the asymmetry
of passage with respect to $\Delta$, in our case could be affected by
drag-effect which distinguishes the fast processes from the slow
ones; the particle on the right well sees the barrier
height to decrease faster (for example for $t<T_1$ during the
very first cycle of $h(t)$) than it sees the height increase (for $t>T_1$
in the same cycle) for $\Delta <0$ and vice versa.
This is the result we wanted to confirm.

	The numerical experiment performed is computationally
expensive even for the two-well system and it is beyond our
means to go for extended periodic systems. However, we make
some extrapolatory remarks. If we have a spatially periodic
potential field, with reflection symmetry about the maxima when
the sweeping external field $h(t)$ is zero, no asymmetric motion
of the Brownian particle could be expected when a symmetric
sweeping field with $(\Delta =0)$ is applied. Let the minima of
the potential wells $i$ be at $\bar m_i$. Now, we make further
idealization to make contact with our experiment: A Brownian
particle at the $i$th well
feels as though it were in the left well of the potential $\Phi
(m)$ for its motion towards the $(i+1)$th well and similarly
for its motion towards the $(i-1)$th well it feels as though it
were in the right well of $\Phi (m)$. Now, $h(t)>0 (<0)$ would
make the common tangent to all the
minima tilt with negative(positive) slope. Thus, a symmetric saw
tooth field $h(t)$ $(\Delta=0)$ will make the common tangent
change its slope with time at the same uniform rate on either
direction (negatine
slope to positive slope and vice versa). However, $\Delta
\not=0$ makes the rate of flapping ($-$slope to $+$slope)
different from ($+$slope to $-$slope), reminiscent of the
flagellar strokes of sperm molecules of eukaryotes. (The analogy
is, however, not to be taken literally for in our case the three
important time scales involved, namely, the relaxation time of
the local minima, the $({\dot h \over h_{c}})^{-1}$ and the mean first
passage time $<\tau >$, are required to be comparable.[8]) Given
the asymmetric nature of the result we have obtained for the two
well system, it is plausible that owing to the application of a
zero-time-averaged asymmetric field a preferentially asymmetric
motion of the Brownian particle in the periodic system potential
field will ensue.[7]

	In summary, we perform a (Langevin dynamic) numerical
experiment on a two-well system subjected to Gaussian white
noise (fluctuating) force and driven by an external saw-tooth
type field and show that the passages of a Brownian
particle from one well to the other have a different behaviour
than the passages in the reverse direction if the driving field
has temporal asymmetry. The conclusion is based on the study of
hysteresis loop area calculated for the two-well system. It is
important to note that in our calculation, unlike in Ref.6,
noise plays a crucial role for the applied field amplitude $h_0$
is restricted to a value lower than the critical value $h_c$ and
hence making the passages impossible without noise. We
extrapolate to conclude that asymmetric passage or transport is
possible in a symmetric (nonratchet-like) periodic potential
provided the system is driven by temporally asymmetric driving
force (Fig.1) and assisted by (additive) Gaussian white noise.
Moreover, our numerical experiment is amenable to real
experiment such as the one performed recently by Simon and
Libchaber[9] on optical systems with double-well potential.

\vfill
\eject
\noindent{\bf References}
\vskip .3 true cm

\item{[1]} M.O. Magnasco, {\it Phys. Rev. Lett.} 71 (1993) 1477,
and references therein.
\item{[2]} R. Bartussek, P. Hanggi, and J.G. Kissner, {\it
Europhys. Lett.} 28 (1994) 459.
\item{[3]} C.R. Doering, W. Horsthemke, and J. Riordan, {\it Phys.
Rev. Lett.} 72 (1994) 2984; L.P. Faucheux, L.S. Bourdieu, P.D.
Kaplan, and A.J. Libchaber, {\it Phys. Rev. Lett.} 74 (1995) 1504.
\item{[4]} M. Bier and R.D. Astumian, {\it Phys. Rev. Lett.} 71
(1993) 1649; R.D. Astumian and M. Bier, {\it Phys. Rev. Lett.} 72
(1994) 1766.
\item{[5]} M.M. Millonas and M.I. Dykman, {\it Phys. Lett.} A185
   (1994) 65; M.M. Millonas, {\it Phys. Rev. Lett.} 74 (1995) 10;
   A.M. Jayannavar, {\it Phys. Rev.} E (submitted).
\item{[6]} A. Ajdari, D. Mukamel, L. Peliti, and J. Prost, {\it
Journ. Phys. 1 France} 4 (1994) 1551.
\item{[7]} D.R. Chialvo, and M.M. Millonas, {\it preprint};
M.M. Millonas, and D.R. Chialvo, {\it Phys. Rev. Lett.}
(submitted).
\item{[8]} M.C. Mahato and S.R. Shenoy, {\it J. Stat. Phys.} 73
(1993) 123; {\it Phys. Rev.} E50 (1994) 2503, and references
therein.
\item{[9]} A. Simon, and A. Libchaber, {\it Phys. Rev. Lett.} 68
(1992) 3375.
\vfill
\eject
\noindent{\bf Figure Captions}
\vskip .3 true cm
\item{Fig. 1.} Depicts the external periodic field $h(t)$. We have
taken amplitude $h_0$=0.7$h_c$, and period $T_0$=28.0.
\item{Fig. 2.} Part of the first-passage time distribution $\rho
(\tau)$ for the field sweep shown in Fig.1. $\rho(\tau)$ extends
to 24 cycles of $h(t)$ in 15000 runs for $D=0.5$ and $h_0$=0.7$h_c$.
\item{Fig. 3.} Hysteresis loop area versus the asymmetry $\Delta
= {(T_1 -T_0/2)\over {T_0/2}}$ in the $\pm$slopes of $h(t)$
(Fig. 1) for $D=0.3(\odot)$ and $D=0.5(\otimes)$. For each data
point $\rho (\tau)$ calculated from a minimum of 10000
first-passage-times $\tau$ from the right well to the left well were
used. The vertical dashed line at $\Delta =0$ is drawn only for
convenience.
\vfill
\eject
\end